\documentclass[aps,prl,preprint,groupedaddress]{revtex4}

\usepackage{graphics}
\usepackage{graphicx}

\def\yp {{y^+}}
\def\utau {{u_\tau}}
\def\Retau {{Re_\tau}}

\def\lz  {{\lambda_z  }}
\def\lx  {{\lambda_x  }}
\def\lzp {{\lambda_z^+}}
\def\lxp {{\lambda_x^+}}

\def\reffig#1{Fig.~\ref{fig:#1}}

\begin{document}
\title{On a self-sustained process at large scale in the turbulent channel flow}
\author{Yongyun Hwang}
\email{yongyun@ladhyx.polytechnique.fr}
\affiliation{LadHyX, CNRS -- \'Ecole Polytechnique, 91128 Palaiseau, France}
\author{Carlo Cossu}
\email{carlo.cossu@imft.fr}
\affiliation{ IMFT -- CNRS , All\'ee du Professeur Camille Soula, 31400 Toulouse, France}
\affiliation{D\'epartement de M\'ecanique, \'Ecole Polytechnique, 91128 Palaiseau, France}

\begin{abstract}

Large-scale motions, important in turbulent shear flows, are frequently attributed to the interaction of structures at smaller scale.
Here we show that, in a turbulent channel at $Re_\tau \approx 550$, large-scale motions can self-sustain even when smaller-scale structures populating the near-wall and logarithmic regions are artificially quenched.
This large-scale self-sustained mechanism is not active in periodic boxes of width smaller than $L_z \approx 1.5 h$ or length shorter than $L_x \approx 3 h$ which correspond well to the most energetic large scales observed in the turbulent channel.

\end{abstract}

\maketitle


Understanding the mechanisms by which turbulence self-sustains in wall bounded turbulent shear flows is still an open challenge that is relevant for applications ranging from atmospheric dynamics to the design of low drag vehicles.
One of the most robust features of these flows is the presence of streamwise streaks, i.e. narrow streamwise regions where the streamwise velocity is larger or smaller than the mean velocity at the same distance from the wall.
The existence of streaks in the buffer layer with a characteristic average spanwise scale $\lambda_z^+ \approx 100$, in wall units,
has been revealed by early experimental observations \cite{Kline1967,Smith1983}. Numerical experiments furthermore reveal that turbulence is not sustained in periodic domains smaller than this spacing \cite{Jimenez1991}.
There is now large consensus that these near-wall structures are generated via a self-sustained process \cite{Hamilton1995,Waleffe1998} based on the streaks amplification from the vortices via the lift-up effect \cite{Landahl1990} followed by the breakdown of the streaks \cite{Waleffe1995,Schoppa2002} and the regeneration of the vortices.
This buffer-layer process survives when turbulent structures are artificially removed from the outer layers \cite{Jimenez1999}.

Coherent structures, however, exist at larger scales. It has long been known that streamwise velocity correlations are important up to lengths of the order of the outer length scale $h$ of the flow (e.g. the channel half-width, the pipe radius or the boundary layer thickness), and have been related to the presence of `bulges' or `large-scale motions' \cite{Kovasznay1970}.
Recently, it has been also realized that coherent streaks exist at even larger scale with typical spanwise and streamwise scales $\lz \approx O(h)$ and $\lx \approx O(10h)$ respectively \cite{Jimenez1998,Tomkins2003,Guala2006,Hutchins2007,Balakumar2007}.
These very large-scale streaks, also referred to as very large-scale motions or global modes or superstructures, are important because they carry a very significant fraction of the turbulent kinetic energy and of the turbulent Reynolds stress \cite{Kim1999,Tomkins2003,Guala2006,Balakumar2007} contradicting the early view that the motions at very large scale are essentially inactive \cite{Townsend1976}.

The fundamental question in which we are interested here is: What is the origin of the streaky motions at very large scale?
It has been conjectured that these motions could result from the concatenation of large-scale motions \cite{Kim1999,Guala2006}.
These large-scale motions are in their turn made of the aggregation of a huge number of hairpin vortices of smaller scale \cite{Tomkins2003,Wu2009}.
Even if from a different perspective, further numerical experiments \cite{Toh2005} support the idea that very large scale motions can be directly forced and interact with near-wall structures, at much smaller scale, via a co-supporting cycle. In this mechanism, while the near-wall cycle would continue to exist even in the absence of large scale structures, the reverse would not happen.

The current wisdom is therefore that large- and very large-scale motions would not exist in the absence of smaller scale structures.
An alternative way of thought is however emerging from recent findings that extend recent hydrodynamic stability methodologies to fully developed turbulent flows.
It has recently been theoretically predicted that very large scale streamwise streaks can be amplified by a coherent lift-up effect which is able to extract energy from the mean flow at very large scale, without the mediation of coherent structures at smaller scale \cite{delAlamo2006,Pujals2009,Cossu2009,Hwang2010,Hwang2010b}.
The existence of the coherent lift-up has been confirmed in experiments where very large scale coherent streaks have been artificially induced by forcing coherent streamwise vortices \cite{Kitoh2008,Pujals2009b,Pujals2010b}.
It is therefore tempting to conjecture that the very large scale motions could result from a self-sustained mechanism bearing some similarity to the processes observed in the buffer-layer and in transitional flows (see e.g. the discussions in \cite{Guala2006,Jimenez2007,Cossu2009}). The goal of the present study is to verify if this conjecture is true.

We proceed by designing a `conceptual' numerical experiment where we gradually remove the smaller scales from the flow while still taking into account the associated dissipation.
The ideal tool to perform such a kind of experiment is the large eddy simulation (LES), which has also the advantage of keeping manageable the computations in very large domains at high Reynolds numbers.
In LES the very small scale motions are not resolved by the grid and the residual stress tensor associated with these small scale filtered motions is modelled with an eddy viscosity $ \tau_{ij}-\tau_{kk}\delta_{ij}/3=-2\nu_T \overline{S}_{ij}$, where  $\overline{S}_{ij}$ is the symmetric part of the filtered velocity gradient.
From the adopted perspective it is important to avoid any backward spectral energy transfer from the smaller to the larger scales. Therefore, we consider the `static' Smagorinsky model (Smagorinsky, 1963) where the eddy viscosity is simply modeled as $\nu_T =D (C_S \overline{\Delta})^2\overline{\mathcal{S}}$ where, $\overline{\mathcal{S}}=(2\overline{S}_{ij}\overline{S}_{ij})^{1/2}$ is a measure of the local shear and $C_S \overline{\Delta}$ is the local mixing length, given by the product of the filtering scale $\overline{\Delta}$ to the
Smagorinsky constant $C_S$. The van Driest damping function $D=1-e^{-(y^+/A^+)^3}$ is also used to enforce a physical behavior near the wall (for the further details on properties of the Smagorinsky model, the reader may refer e.g. to \cite{Hartel1998,Meneveau2000}).
Dynamic Smagorinsky models, that are known to perform better for engineering applications, are not used here because we want to avoid backscatter.
The simulations are performed using the public domain code {\tt diablo}  \cite{Bewley2001} which implements the fractional-step method based on a semi-implicit time integration scheme and a mixed finite-difference and Fourier discretization in space.
The computational domain is chosen large enough ($L_z=6h$) to contain three to four large scale streaks in the spanwise direction and long enough ($L_x= 56 h$) to contain the longest observed streamwise wavelengths of the very large scale motions.
The grid size of $384 \times 65 \times 96$ points in respectively the streamwise ($x$), wall-normal ($y$) and spanwise ($z$) direction is set-up so as to resolve the near-wall cycle properly in the reference simulations \cite{Zang1991}. A grid stretching is applied in the wall-normal direction with the first grid point at $y^+=1.8$ and a maximum spacing $\Delta_y^+=40$ at the channel center.

\begin{figure}
\begin{center}
\includegraphics[width=0.49\textwidth]{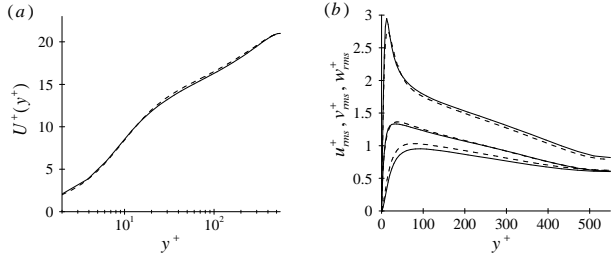}
\end{center}
\vspace*{-3mm}
   \caption{\label{fig:BaseLineMeanVel}
   Comparison of mean statistics of the reference case ($C_S=0.05$) with data from the direct numerical simulations in Ref.~\cite{delAlamo2003} at the same $\Retau=550$:
   (a) mean streamwise velocity profile $U(\yp)$; (b) turbulence intensity profiles $u_{rms}(\yp)$, $v_{rms}(\yp)$ and $w_{rms}(\yp)$.}
\end{figure}

\begin{figure}
\includegraphics[width=0.45\textwidth]{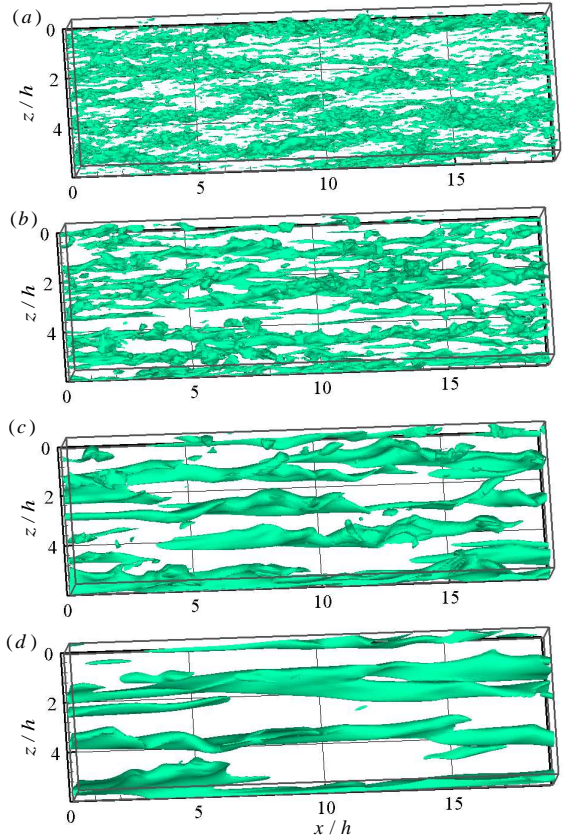}
   \caption{\label{fig:Streaks} (Color online).
   Isosurface ($u^+=-2$) of the instantaneous streamwise velocity fluctuation for:
   the reference simulation with $C_S=0.05$ (a) and for simulations performed with
   increasing artificial dissipation of the small-scales with
   $C_S=0.1$ (b), $C_S=0.2$ (c) and $C_S=0.3$ (d).
   }
\end{figure}

\begin{figure}
\includegraphics[width=0.5\textwidth]{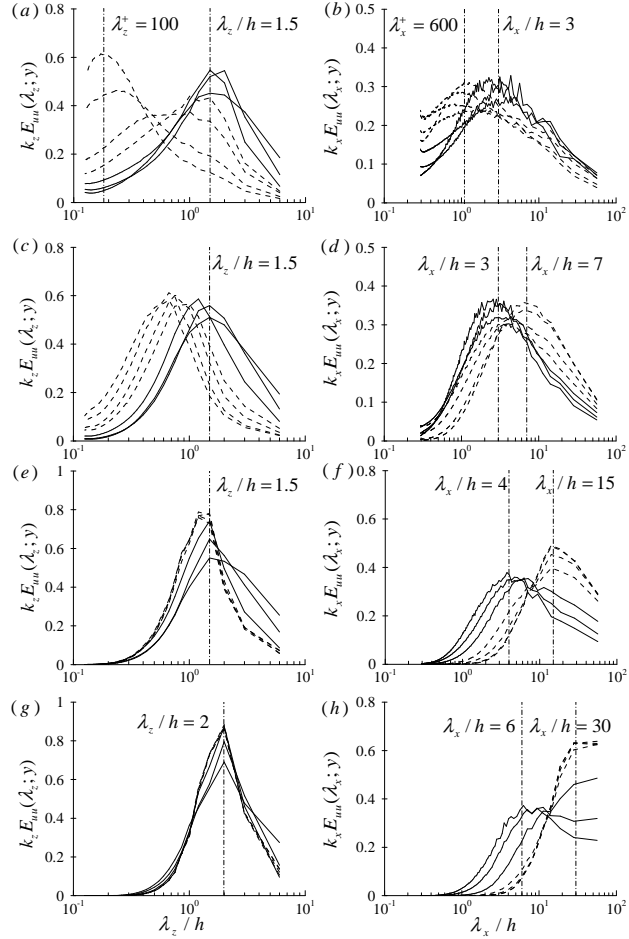}
   \caption{\label{fig:PremSpec}
   Spanwise premultiplied power spectrum   $k_z E_{uu} (\lz)$ [(a), (c), (e) and (g)] and
   streamwise premultiplied power spectrum $k_x E_{uu} (\lx)$ [(b), (d), (f) and (h)]
   for respectively the reference simulation with
   $C_S=0.05$ [(a) and (b)]  and for the cases
   $C_S=0.1 $ [(c) and (d)],
   $C_S=0.2 $ [(e) and (f)] and
   $C_S=0.3 $ [(g) and (h)].
   The premultiplied spectra are extracted in the inner- ($y^+=16, 30, 70, 108$ dashed lines)
   and in the outer layer ($y/h=0.38, 0.65, 1$ solid lines). 
      }
\end{figure}

The simulations are performed by enforcing a constant mass flux corresponding to the constant $Re_m=20133$, where $Re_m=U_m 2h /\nu$ is the Reynolds number based on the bulk velocity $U_m=(1/2h) \int_{-h}^h U~ dy$, the kinematic viscosity $\nu$ and the channel half-height $h$. The reference case is first considered by choosing the Smagorinsky constant $C_S=0.05$ which was shown to provide the best accordance with statistics of direct numerical simulations \cite{Hartel1998}.
The reference solution exhibits an average friction-based Reynolds number $\Retau=550$, where $\Retau=h \utau / \nu$ and $\utau$ is the friction velocity.
The mean streamwise velocity profile and the turbulence intensities agree well with the those issued from direct numerical simulations at the same $\Retau$ \cite{delAlamo2003} as reported in \reffig{BaseLineMeanVel}.
A snapshot of the streamwise velocity fluctuation in the reference case is reported in \reffig{Streaks}(a). The structure of this field is very similar to the one obtained from direct numerical simulations, with large-scale streaks embedded in clouds of smaller scale structures (see e.g. \cite{Jimenez2007}). The premultiplied spanwise and streamwise velocity spectra $k_z E_{uu} (k_z)$ and $k_x E_{uu} (k_x)$ corresponding to the reference case, reported in \reffig{PremSpec}(a) and \ref{fig:PremSpec}(c) respectively, are also in qualitatively good accordance with those issued from DNS at similar $\Retau$ \cite{Jimenez1998}.
In particular, in the premultiplied spectra, the scales of the buffer-layer cycle  $\lzp \approx 100$ and $\lxp \approx 600$ are observed for $y^+<30$ , while in the outer region the peaks corresponding to the large-scale motions are observed at $\lz/h\approx 1.5\sim 2$ and $\lx/h\approx 3\sim 4$ with long tails in the range of $\lz/h> 3\sim 4$ that represent the very-large-scale motions \cite{Kim1999,delAlamo2004,Monty2009}.

To test if the large-scale structures can survive even in the absence of smaller scales we progressively damp the small-scale structures by simply repeating the simulations with larger values of the Smagorinsky constant $C_S$ which only affects the diffusion term and not the nonlinear advective terms.
In these additional simulations the Reynolds number is kept at its reference value $Re_m=20133$. The friction-based Reynolds number is not strongly affected by the change in $C_S$: $\Retau=494$ is found for $C_S=0.1$,  $\Retau=518$ for $C_S=0.2$ and $\Retau=560$ for $C_S=0.3$.

For increasing $C_S$ values, the small-scale coherent motions are increasingly damped, as can be seen from the instantaneous streamwise velocity fluctuation fields reported in \reffig{Streaks}
Only large- and very large-scale streaky motions survive for $C_S \approx 0.3$ (\reffig{Streaks}$d$) and also these motions are quenched for $C_S=0.4$.
The effects observed on the instantaneous velocity fields are confirmed by the analysis of the premultiplied spectra reported in \reffig{PremSpec}.
For $C_S=0.1$, the buffer-layer peak $\lz^+\approx 100$ in the spanwise premultiplied spectrum $k_z E_{uu}(\lz)$ has been quenched, meaning that the near wall cycle has been suppressed. However, structures with almost constant $\lz/y$ survive in the log-layer for $C_S=0.1$, as seen in \reffig{PremSpec}(c). For larger $C_S$, also these structures are quenched. Only one peak survives with $\lz \approx 1.5 h$ for $C_S=0.2$ and $\lz =2 h$ for $C_S=0.3$ with $\lz$ corresponding well to the scale of large- and very-large scale motions ($\lz \approx 1.5 - 2h$) apparent in \reffig{PremSpec}$a$ and observed in numerical simulations and experiments \cite{Jimenez1998,Balakumar2007}.

The analysis of the streamwise  premultiplied spectra $k_x E_{uu} (\lx)$, confirms this scenario.
For $C_S=0.1$ the $\lx^+ \approx 600$ peak has been replaced by much longer structures ($\lx \approx 7 h$).
In the channel center, structures with $\lx \approx 3h$ dominate like in the reference simulation.
The typical streamwise scales in the inner layer, no more hidden by the near-wall cycle, are now longer than in the outer layer.
These scales grow longer when $C_S$ is further increased (see \reffig{PremSpec}(d),(f),(h)).
The streamwise scales surviving for $C_S=0.3$ are larger by a factor of two when compared to the `natural' large-scale and very large-scale motions, which can probably be attributed to the very important artificial dissipation enforced at this $C_S$.
The analysis of the wall-normal structure of the modes associated with these scales, reveals that the modes associated to longer $\lx$, when compared to shorter $\lx$ ones: (a) have turbulence intensity peaks nearer to the wall, (b) penetrate more to the wall preserving a larger turbulence intensity, (c) have a ratio of the wall-normal to the streamwise turbulence intensity which is smaller.
These structures remain essentially outer-layer structures even for the longest $\lx$.

These findings, especially for the case $C_S=0.1$, are in accordance with recent numerical experiments where it was shown that large scale motions in the outer layers are almost independent of the buffer-layer processes \cite{Flores2006,Flores2007}.
Even more importantly, these results prove that large-  and very-large scale motions in the turbulent channel do not necessarily rely on the existence of smaller scale structures like buffer-layer streaks and vortices or hairpin vortices.
This is a strong indication that an independent self-sustained process at large scale exists in the turbulent channel and probably in the other turbulent canonical shear flows such as the pipe and the boundary layer, as was previously conjectured \cite{delAlamo2006,Guala2006,Cossu2009}.

The self-sustained process at large scale is associated with coherent large scale streaks that undergo sinuous oscillations and breakdown in an apparently random way. This phenomenology is similar to what is observed in minimal box realizations of the buffer layer self-sustained process and in the subcritical laminar-turbulent transition in the channel \cite{Jimenez1991,Hamilton1995,Waleffe1995,Reddy1998}.
In these situations, the process can self-sustain only in periodic boxes sufficiently large for the streaks to grow by lift-up effect and sufficiently long to allow the sinuous streaks oscillations to develop.
We have therefore repeated the simulations for $C_S=0.3$ in smaller horizontally-periodic boxes. A constant spatial resolution $(\Delta x, \Delta z$) is used so as to maintain the LES filter width, unchanged.
A detailed analysis of numerous box combinations reveals that the large-scale process is sustained only if the streamwise and spanwise box sizes are larger than the minimal values $L_{x,min} \approx 3 h$ and $L_{z,min} \approx 1.5 h$ respectively.
These values are in very good agreement with the spatial length-scales of most energetic outer layer motions \cite{Jimenez1998,delAlamo2004,delAlamo2006,Balakumar2007}.
This similarity of the minimal box sizes to the characteristic scales of the most energetic structures observed in the `natural' turbulent flows is also observed for buffer-layer structures \cite{Jimenez1991},
The minimal spanwise size $L_{z,min}$ is also well in the range of scales where streaks are strongly amplified by the coherent lift-up effect in turbulent channels \cite{delAlamo2006,Pujals2009}.

The analysis of the power input and dissipation $I(t)$ and $D(t)$ in the minimal box $L_{x,min} \times L_{z,min}$ for $C_S=0.3$
reveals pseudo-periodic oscillations with period $\approx 13-15 h/U_b$ associated  with sinuous oscillations of the low speed streaks in the whole channel. These oscillations are typically followed, in an intermittent way,  by a large excursion in the drag and the dissipation that is very similar to the `bursting' observed in the buffer-layer process \cite{Jimenez1991}.

From the minimal box results we also conclude that the existence of motions at very large scale ($\lx > 3h$) is not necessary for the survival of the motions at large scale ($\lx \approx 3h$).
Whether the motions with $\lx > 3h$ are active or are passively forced by the  process at $\lx \approx  3h$, is still an open question that is under current scrutiny.

\acknowledgements{
We gratefully acknowledge the use of the {\tt diablo} code mainly developed by T.R.~Bewley and J.~Taylor, partial support from DGA and \'Ecole Polytechnique and NORDITA where part of this work was done.
}


\newcommand{\noopsort}[1]{} \newcommand{\printfirst}[2]{#1}
  \newcommand{\singleletter}[1]{#1} \newcommand{\switchargs}[2]{#2#1}

\end{document}